\documentclass[conference]{IEEEtran}
\IEEEoverridecommandlockouts
\usepackage{cite}
\usepackage{amsmath,amssymb,amsfonts}
\usepackage{algorithmic}
\usepackage{graphicx}
\usepackage{textcomp}
\usepackage{xcolor}
\usepackage{mathptmx}
\usepackage{color}
\usepackage{xspace}
\usepackage{url}

\newcommand{\toolNameF}{\emph{Continuous Learning Platform}\xspace}
\newcommand{\toolNameA}{\emph{CLP}\xspace}

\newcommand{\collection}{\emph{Data Collection}\xspace}
\newcommand{\collectionDSLoader}{\emph{Dataset Loader}\xspace}
\newcommand{\collectionDLoader}{\emph{Driver Loader}\xspace}
\newcommand{\collectionImporter}{\emph{Importer}\xspace}

\newcommand{\management}{\emph{Data Management}\xspace}
\newcommand{\managementDA}{\emph{Data Aligner}\xspace}
\newcommand{\managementLC}{\emph{Label Consolidator}\xspace}

\newcommand{\managementDCo}{\emph{Data Composer}\xspace}

\newcommand{\storage}{\emph{Unified Dataset}\xspace}

\newcommand{\distribution}{\emph{Data Distribution}\xspace}
\newcommand{\distributionCB}{\emph{Classifier Builder}\xspace}
\newcommand{\distributionOC}{\emph{Online Classifier}\xspace}
\newcommand{\distributionCD}{\emph{Classifier Deployer}\xspace}

\def\BibTeX{{\rm B\kern-.05em{\sc i\kern-.025em b}\kern-.08em
    T\kern-.1667em\lower.7ex\hbox{E}\kern-.125emX}}
\begin{document}

\title{A Platform to Collect, Unify, and Distribute Inertial Labeled Signals for Human Activity Recognition}

\author{
\IEEEauthorblockN{Anna Ferrari, Daniela Micucci, Marco Mobilio, and Paolo Napoletano}
\IEEEauthorblockA{Department of Informatics, Systems and Communication, %\\
University of Milano Bicocca\\
Email: a.ferrari34@campus.unimib.it, {daniela.micucci|marco.mobilio|paolo.napoletano}@unimib.it}
%\and
%\IEEEauthorblockN{Daniela Micucci}
%\IEEEauthorblockA{Department of Informatics, Systems and Communication\\
%University of Milano Bicocca\\
%Email: daniela.micucci@unimib.it}
%\and
%\IEEEauthorblockN{Marco Mobilio}
%\IEEEauthorblockA{Department of Informatics, Systems and Communication\\
%University of Milano Bicocca\\
%Email: marco.mobilio@unimib.it}
%\and
%\IEEEauthorblockN{Paolo Napoletano}
%\IEEEauthorblockA{Department of Informatics, Systems and Communication\\
%University of Milano Bicocca\\
%Email: paolo.napoletano@unimib.it}

}

\maketitle

\begin{abstract}
Human activity recognition (HAR) is a very active research field. Recently, deep learning techniques are being exploited to recognize human activities from inertial signals. However, to compute accurate and reliable deep learning models, a huge amount of data is required. %Moreover, the creation of a dataset to be used with deep learning techniques is an onerous process that requires the involvement of a significant number of possibly heterogeneous subjects. 
%The publicly available datasets are few and, with rare exceptions, contain few subjects. Furthermore, datasets are heterogeneous and therefore not directly usable all together. 
The goal of our work is the definition of a platform to support long-term data collection to be used in training of HAR algorithms. The platform aims to integrate datasets of inertial signals in order to make available to the scientific community a large dataset of homogeneous data. % and, when possible, enrich it with context information (e.g., characteristics of the subject, device position, and so on). 
%Moreover, the platform has been designed to provide additional services such as the deployment of activity recognition models and online signal labelling services. 
The architecture of the platform has been defined and some of the main components have been developed in order to verify the soundness of the approach.
\end{abstract}

% no keyworwds

\IEEEpeerreviewmaketitle

\section{Introduction} \label{sec:intro}
Recognition of activities of daily living (ADL) from inertial signals is a very active research field in view of the many application domains interested in (e.g., sports~\cite{survey_energy} and healthcare~\cite{survey_Park}) and the increasing diffusion of wearable devices embodying inertial sensors.
However, the classification of sensor data %with respect to the actions performed by humans 
represents the main challenge of human activity recognition (HAR) because the space of the signals are not perfectly separated. 

Preliminary HAR techniques exploited supervised machine learning algorithms. 
%Supervised machine learning however presents several challenges:
%
The main challenges of these techniques include: the difficulty of transferring the performances achieved in laboratory to a real context~\cite{bagala2012evaluation}, and the inability of the algorithms to extract and organize discriminative information from the data~\cite{bengio2013deep}. % due to the strong dependency between the selection of the features and the performance of the classification~\cite{bengio2013deep}. 

%lecun1995convolutional,
In recent years, deep learning has been successfully applied to 3D and 4D signals, and more recently it has been exploited also for 1D signals~\cite{ronao2016human,ferrari2019ISCT}. The widespread use of deep learning techniques is justified by their properties of \textsl{local dependency} and \textsl{scale invariance}~\cite{zeng2014convolutional}. Furthermore, deep learning methodologies permit automated discovery
of abstraction which overcomes the features extraction issue~\cite{bengio2013deep}. While deep learning techniques are powerful and achieve high performance, they rely on very complex models that strictly depend on the estimation of a large number of parameters, which requires a considerable amount of available data~\cite{bianco2018benchmark}.

In recent years, some researchers have published their own dataset related to HAR. However, these datasets are heterogeneous and their standardization in a single unified dataset requires considerable effort. For example, signals are expressed in different units of measure, they may include gravity or not, and signals have different acquisition frequencies. Furthermore, labels are not aligned with a common dictionary and sometimes have different meanings among different datasets ('sitting' may refer to the state of being seated in a chair or the transition from standing to sitting).

Since acquiring labeled time series is a costly procedure in terms of resource, time, and people involved, we think that integrating existing datasets is the right direction despite the strong heterogeneity of the data. This abstract presents a platform that semi-automatically integrates heterogeneous data and provides them in a homogenous form. 

\section{\toolNameF} \label{platform}
The main aim of the \toolNameF (\toolNameA in the sequel) is to make available (i) a large amount of labeled inertial signals related of ADLs and falls, (ii) a catalogue of downloadable activity recognition models, and (iii) a service that, given a set of raw data, identifies the corresponding ADL. 
 
\begin{figure}[htb]
  \centering
  \includegraphics [width = 0.5\textwidth]{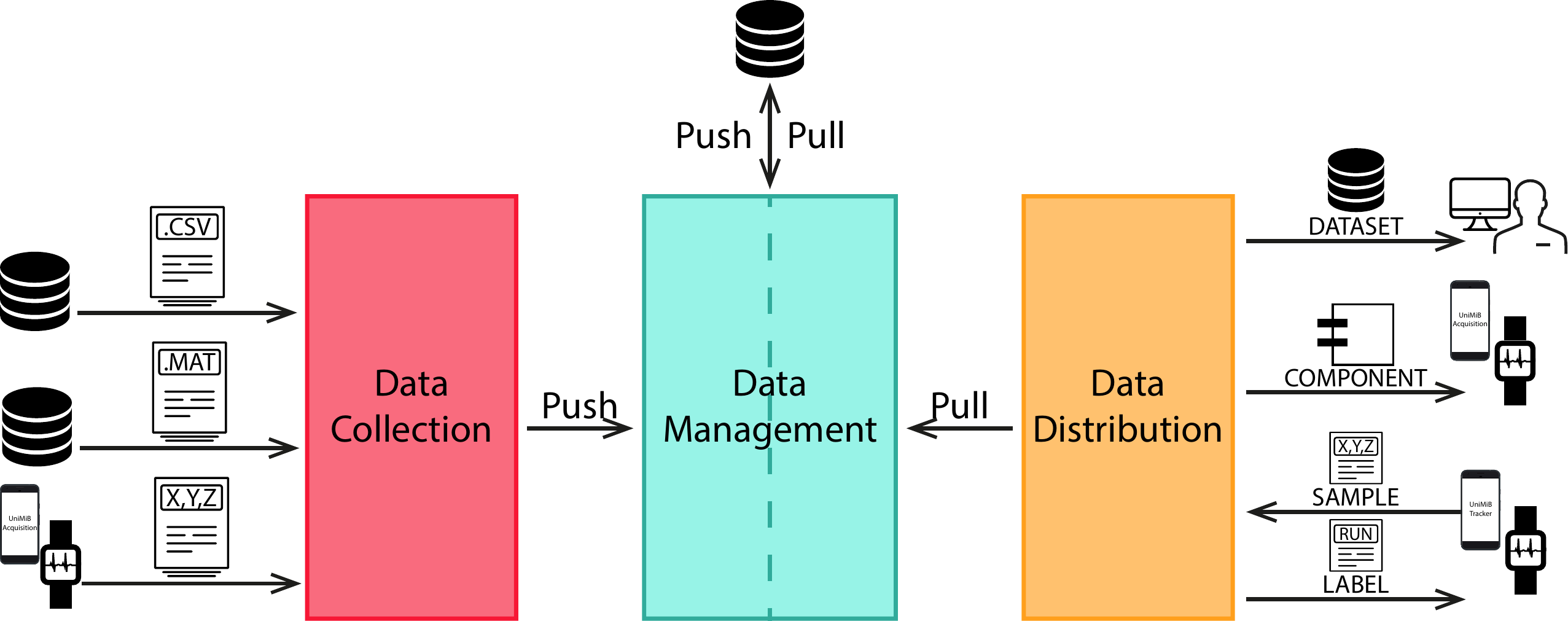}
  \caption{Overview of the platform.}
  \label{fig:overview}
\end{figure}

\toolNameA \emph{collects} inertial signals from existing datasets or applications, \emph{manages} the collected inertial signals, and \emph{distributes} uniformed labeled inertial signals, activity recognition models, and labels assigned to series of inertial signals (Figure~\ref{fig:overview}). %Those functionalities are respectively reified by the \collection, \management, and \distribution components. Figure~\ref{fig:overview} sketches the overall architecture. %and its interaction with the external actors.

%The following sections provide detailed information about how the components have been architecturally defined and about our preliminary development results.

\subsection{\collection} \label{collection}
The \collection component acquires existing inertial signals and uniforms their representations. Existing inertial signals can be from i) existing datasets (e.g., UniMiB SHAR~\cite{app7101101}); ii) labeled inertial signals from applications used by volunteers to record datasets (e.g., UniMiB AAL~\cite{AAL}); and iii) non-labeled inertial signals from applications used by people performing ADLs (e.g., Sensor Data Logger~\cite{sensordata}). 
%Figure~\ref{fig:collection} sketches the modules we identified for the \collection component. 
%
%\begin{figure}[htb]
%  \centering
%  \includegraphics [width = 0.4\textwidth]{Images/Collection.pdf}
%  \caption{The \collection component.}
%  \label{fig:collection}
%\end{figure}

One of the main issues in handling multiple datasets is the lack of consistency in terms of \emph{how} the data is stored and \emph{what} are the information (e.g., in the UCI HAR dataset~\cite{anguita2013public} data are organized in txt files stored in 2 directories; in MobiAct~\cite{vavoulas2016mobiact} data are organized in csv files stored in 20 directories). %For example, in the UCI HAR dataset~\cite{anguita2013public} data are stored in 2 directories (train and test) which contain .txt files; in MobiAct~\cite{vavoulas2016mobiact} data are subdivided in 20 directories (one per each activity) which include .csv files. 
In \toolNameF we enforce a single storage technology for all data and a single structure for the data. 

The \collection component includes the \collectionDLoader, the \collectionDSLoader, and the \collectionImporter modules, which respectively allow to load custom drivers developed to support specific datasets, to load datasets to be integrated in \storage, and to ask for the integration of the new datasets into the \storage. Separating the \collectionDSLoader from the \collectionDLoader, allows the dynamic on-boarding of the driver, which may require a reboot of the \collectionDSLoader service in order to be visible and exploitable from the service itself. 

\subsection{\management} \label{management}
The \management component integrates the new labelled signals into the \storage and makes available sets of labelled signals to those who need them (\emph{Data Access}). 
Before being included into the \storage, signals require to be homogeneous both in terms of \emph{representation} and \emph{label}. Homogeneous representations of the input data are required by machine learning methods~\cite{baltruvsaitis2019multimodal}. Labels must be unique for the same activity and the same labels must be assigned to the same signals, otherwise the classifiers will not work efficiently. %Signals are required to be  Signals from different datasets may have assigned different labels for the same ADL (e.g., 'walk' vs 'walking'), same label for different ADL (e.g., 'sitting' may refer to the state of being seated or the transition from standing to sitting), and different labels for the same activity (e.g., 'running' vs 'jogging').

%%Figure~\ref{fig:management}.
%
%\begin{figure}[htb]
%  \centering
%  \includegraphics [width = 0.5\textwidth]{Images/Management.pdf}
%  \caption{The \management component.}
%  \label{fig:management}
%\end{figure}

The \management component includes the following modules. % (Figure~\ref{fig:management}).
The \managementDA module pre-processes the data from the \collection component in order to make them usable by any machine learning method. For example, an activity that is in charge of the \managementDA module is the conversion to a same measurement unit.
The \managementLC module is in charge of uniforming the labels of the dataset to include to a common unified set. 
For example, if a dataset uses the label 'sitting' to label signals related to the transition (from standing to sitting down) and in the \storage is used 'sit down' to label signals related to the transition (from standing to sitting down), then the label will be changed to be consistent with the \storage. In view of the delicate nature of this procedure, this module is intended to be semi-automatic: it provides suggestions on the assignment of labels, but ultimately it is down to the end user to decide whether or not to accept the suggestions. 
%
%The \managementDC module is in charge of assigning labels to inertial signals that have not been labeled. These types of signals may result from applications that acquire signals only without providing any classification. This module can exploit an activity recognition model already trained.
%
In terms of data distribution, the \managementDCo module simply intercepts requests for sets of labeled signals. For example, a request can be: ``all signals labeled \emph{running}''.

\subsection{\distribution}\label{distribution}
The \distribution component role is to provide i) sets of labeled signals according to specific needs; ii) trained classifiers; and iii) labels corresponding to the activities performed given frames of signals.
%real-time classification frames of signals.

%The availability of sets of labeled signals helps researchers that need to validate their techniques with a wide and public data set. The huge amount of data also allows to work with deep learning techniques, and publicly available datasets allows to generalize the results and to compare techniques.

%Figure~\ref{fig:distribution} sketches the \distribution component architectural organization.
%
%\begin{figure}[htb]
%  \centering
%  \includegraphics [width = 0.9\textwidth]{Images/Distribution.pdf}
%  \caption{The \distribution component.}
%  \label{fig:distribution}
%\end{figure}

The \distribution component includes the following modules.
The \distributionCB module is in charge of distributing activity recognition models that can be integrated in domain dependant applications. The module also relies on the \distributionCD to store the new trained activity recognition model to be used for online classification. Finally, the \distributionOC module provides online services related to classification: given a set of inertial signals, it provides information regarding the activity the subject is performing. 

\section{Conclusions}\label{conclusion}
The lack of large datasets penalizes the possibility of exploiting deep learning techniques for human activity recognition. 

We designed a platform which integrates data from heterogenous sources and provides several types of access to the unified data.

The framework has been partially implemented. We have prioritized the development of the most challenging components: \collection and \management~\cite{ferrari2019IEEE}. The modules are web services providing the services through RESTFul APIs. Till now, five datasets have been integrated.

\bibliographystyle{IEEEtran}
\bibliography{references}

\end{document}